\documentclass[draftclsnofoot,onecolumn]{IEEEtran}
\usepackage{pifont}
\usepackage{graphicx}
\usepackage{fancyhdr}
\usepackage{amsmath,amssymb,amstext,amsfonts}
\usepackage{slashbox}
\usepackage{bm}
\usepackage{algorithm}
\usepackage{array}
\usepackage{cite}

\usepackage{subfigure}
\usepackage{balance}
\usepackage{algorithmicx}
\usepackage{algpseudocode}

\usepackage{braket}
\usepackage{tabulary}
\usepackage{diagbox}
\usepackage{amsthm}
\usepackage{color}
\makeatletter
\renewcommand{\maketag@@@}[1]{\hbox{\m@th\normalsize\normalfont#1}}%
\makeatother

\begin{document}
\title{Generalized Kennedy Receivers Enhanced CV-QKD in Turbulent Channels for Endogenous Security of Space-Air-Ground Integrated Network}

\author{Shouye Miao, Renzhi Yuan, Bin Cao, Mufei Zhao, Zhifeng Wang, and Mugen Peng
\thanks{
Shouye Miao, Renzhi Yuan, Bin Cao, and Mugen Peng are with the State Key Laboratory of Networking and Switching Technology, Beijing University of Posts and Telecommunications, Beijing, China; Mufei Zhao is with School of Computer Science and Engineering, Northeastern University, Shenyang, 110819, China; Zhifeng Wang is with School of Electronic and Information Engineering, Tongji University, Shanghai, 201804, China.}
\thanks{Corresponding Author: Renzhi Yuan(renzhi.yuan@bupt.edu.cn)}
\thanks{A journal version of this paper is under peer review process. This work is supported by the National Natural Science Foundation of China under No. 62201075.}}

\maketitle

\begin{abstract}
Endogenous security in next-generation wireless communication systems attracts increasing attentions in recent years. A typical solution to endogenous security problems is the quantum key distribution (QKD), where unconditional security can be achieved thanks to the inherent properties of quantum mechanics. Continuous variable-quantum key distribution (CV-QKD) enjoys high secret key rate (SKR) and good compatibility with existing optical communication infrastructure. Traditional CV-QKD usually employ coherent receivers to detect coherent states, whose detection performance is restricted to the standard quantum limit. In this paper, we employ a generalized Kennedy receiver called CD-Kennedy receiver to enhance the detection performance of coherent states in turbulent channels, where equal-gain combining (EGC) method is used to combine the output of CD-Kennedy receivers. Besides, we derive the SKR of a post-selection based CV-QKD protocol using both CD-Kennedy receiver and homodyne receiver with EGC in turbulent channels. We further propose an equivalent transmittance method to facilitate the calculation of both the bit-error rate (BER) and SKR. Numerical results show that the CD-Kennedy receiver can outperform the homodyne receiver in turbulent channels in terms of both BER and SKR performance. We find that BER and SKR performance advantage of CD-Kennedy receiver over homodyne receiver demonstrate opposite trends as the average transmittance increases, which indicates that two separate system settings should be employed for communication and key distribution purposes. Besides, we also demonstrate that the SKR performance of a CD-Kennedy receiver is much robust than that of a homodyne receiver in turbulent channels.
\end{abstract}


\IEEEpeerreviewmaketitle

\section{Introduction}
\subsection{Background and motivation}
Endogenous security plays a crucial role in space-air-ground integrated network and attracts increasing attentions in next-generation wireless communication systems \cite{jiangxing2021cyberspace,liu20206g,wang2022blockchain,zheng2022dynamic}. A typical solution to endogenous security problems is the quantum key distribution (QKD), which provides unconditional security to physical layer communication links thanks to the inherent properties of quantum mechanics \cite{wang2021analysis,zhang2023adaptive,hamdoun2020information,moreolo2024continuous}. QKD protocols can be divided into two categories according to the types of employed information carriers, i.e., the discrete variable-quantum key distribution (DV-QKD) based on individual photons and the continuous variable-quantum key distribution (CV-QKD) based on Gaussian states. Compared with DV-QKD \cite{gyongyosi2019survey,pirandola2020advances}, CV-QKD \cite{grosshans2002continuous,silberhorn2002continuous,grosshans2003quantum,lorenz2004continuous} enjoys higher secret key rate and better compatibility with current optical communication infrastructure, and therefore, attracted large attentions in recent years \cite{gyongyosi2018low,ghalaii2023continuous,li2024discrete,zhang2024continuous,liu2025otfs,zheng2025free,yao2025continuous}.

Current CV-QKD protocols mainly employ coherent detections to recover the transmitted bits \cite{grosshans2002continuous,silberhorn2002continuous,grosshans2003quantum,lorenz2004continuous,gyongyosi2018low}, whose performance is restricted by the standard quantum limit (SQL) \cite{wittmann2010demonstration}. A type of quantum receiver, generalized Kennedy receiver, was applied to the post-selection based CV-QKD to surpass the SQL and improve the secret key rate (SKR) \cite{wittmann2010demonstration}. It was demonstrated that the generalized Kennedy receiver can increase the SKR in either individual attacks or collective attacks compared with coherent receivers \cite{wittmann2010demonstration,zhao2020security,zhao2021post,zhao2024security}. However, these quantum receiver enhanced CV-QKD protocols are restricted in lossy channel only. In space-air-ground integrated network, the most important and also vulnerable links are the satellite-to-ground links due to the presence of atmospheric turbulence \cite{zhu2002free}. To the best knowledge of the authors, the study of generalized Kennedy receiver enhanced CV-QKD in turbulent channels is still absent.

\subsection{Related works}
The CV-QKD protocol using coherent states was proposed by Grosshans and Grangier \cite{grosshans2002continuous}, where the secret key rate (SKR) was guaranteed by the no-cloning theorem. However, this protocol suffers from the ``3dB limit", i.e., the channel loss cannot be larger than 50\%, and thus cannot be applied in practical implementations. To beat the ``3dB limit", post-selection strategy was combined with the coherent state based CV-QKD protocols \cite{silberhorn2002continuous}. In a post-selection strategy, the legitimate users Alice and Bob can always keep those bits with high effective information and discard the rest. Therefore, Bob can enjoy advantages over potential eavesdropper Eve even in a high path loss channel. Because no entanglement or squeezing was needed, coherent states based CV-QKD protocols enjoy good compatibility with existing optical communication infrastructure and thus have attracted increasing attentions \cite{grosshans2002continuous}.

The coherent receivers are usually used in post-selection based CV-QKD protocols \cite{grosshans2002continuous,silberhorn2002continuous,grosshans2003quantum,lorenz2004continuous,gyongyosi2018low}. However, the detection performance of classical coherent receivers is restricted by the SQL and can only be outperformed by using quantum receivers. Quantum receivers employ quantum detection theory to enhance the error rate performance of discriminating quantum states \cite{burenkov2021practical}. The closed-form solution for the optimal quantum detection of distinguishing any two quantum state was obtained by Helstrom \cite{helstrom1969quantum,helstrom1970quantum}. The first realizable quantum receiver for discriminating two binary phase shift keying (BPSK) modulated coherent states $\{\ket{-\alpha},\ket{\alpha}\}$ is the Kennedy receiver proposed by Kennedy \cite{kennedy1973near}, which consists of a displacement operation $\hat{D}(\beta)$ with $\beta=\alpha$ and an on/off photodetector. We called the quantum receivers based on Kennedy receiver's structure the \textit{generalized Kennedy receivers}. In 1973, Dolinar proposed the Dolinar receiver by controlling the displacement value in a real-time way with a feedback loop \cite{dolinar1973optimum}, which was proved to be an optimal quantum receiving structure for discriminating coherent states.

However, due to the lack of fast processing devices and high-performance photodetectors, the first quantum receiver beating the SQL has not been realized until 2008 by Cook et al \cite{cook2007optical}. After then, the study of generalized Kennedy receiver has attracted increasing attention in recent decades. For example, in 2008, Takeoka et al proposed the optimized displacement receiver (ODR) \cite{wittmann2008demonstration, takeoka2008discrimination}, where the displacement value $\beta$ of $\hat{D}(\beta)$ was optimized to achieve a better performance compared with the Kennedy receiver. In 2010, Wittmann et al demonstrated a generalized Kennedy receiver by replacing the on/off photodetector with a photon-number-resolving detector (a type of photon counter) \cite{wittmann2010demonstration}. In 2014, Becerra et al demonstrated that by decreasing the phase mismatch error of the displacement operation, the detecting signal strength of generalized Kennedy receiver can be extended to practical optical communication scenarios with large photon numbers \cite{becerra2015photon}. In 2018, DiMario et al experimentally demonstrated that the generalized Kennedy receiver can enjoy a robust performance in a noisy environment by combining a high-performance displacement operation and photon-number-resolving detector. In 2020, Yuan et al proposed the optimally displaced threshold detection (ODTD) based generalized Kennedy receiver, where both the displacement and the detection threshold are optimized to improve the detection performance, and theoretically quantified the influence of various types of noise and device imperfection on the ODTD receiver \cite{yuan2020kennedy,yuan2020optimally}. Later in 2020, Yuan et al extended the ODTD to turbulent channels and proposed the conditionally-dynamic based Kennedy (CD-Kennedy) receiver to mitigate the influence of turbulence on the detection performance \cite{yuan2020free}. In 2023, Zhao et al extended the ODTD receiver to the ternary phase shift keying (TPSK) modulation \cite{zhao2023optimally} and future extended the ODTD receiver to the quadrature phase shift keying (QPSK) in 2024 \cite{zhao2024security}.

The combination of generalized Kennedy receiver and the CV-QKD was first proposed in \cite{wittmann2010demonstration}, where a photon-number-resolving detector was used to improve the performance of Kennedy receiver and the detector was applied in a post-selection based CV-QKD protocol. Based on this protocol, Zhao et al studied the secret key rate (SKR) performance of using the generalized Kennedy receiver when a thermal noise channel was considered \cite{zhao2020security,zhao2021post}. Besides, the SKR performance of the post-selection based CV-QKD protocol using a generalized Kennedy receiver for QPSK modulation was also studied in \cite{zhao2024security}. However, current studies \cite{wittmann2010demonstration,zhao2020security,zhao2021post,zhao2024security} on CV-QKD with generalized Kennedy receiver are focusing on lossy channels without turbulence. As we know, the most important and also vulnerable links in space-air-ground integrated network are the satellite-to-ground links, where the atmospheric turbulence cannot be ignored. Therefore, it is meaningful to study the performance of CV-QKD protocol with generalized Kennedy receiver in turbulent channels.

\subsection{Contributions}
In this paper, we focused on the performance of post-selection based CV-QKD enhanced by using CD-Kennedy receivers in the presence of atmospheric turbulence. We first derive the bit-error rate (BER) expression of CD-Kennedy receiver with BPSK modulation in turbulent channels, where a $1\times$ N configuration using equal-gain combining (EGC) method is employed. Then we drive the SKR expression of the binary modulated CV-QKD protocol by using CD-Kennedy receiver with a post-selection on both the detected number of photons and the measured channel transmittance. For comparison, we also derived the corresponding BER expression of homodyne receiver with EGC and the SKR expression of binary modulated CV-QKD protocol by using homodye receiver with a post-selection on both the detected amplitude and the measured channel transmittance. Numerical results demonstrate the BER performance of CD-Kennedy receiver is better than that of homodyne receiver when the average transmittance is large; while the SKR performance of CD-Kennedy receiver is better than that of homodyne receiver when the average transmittance is small. Besides, we also demonstrate that the SKR performance of CD-Kennedy receiver is much robust compared with that of homodyne receiver to the atmospheric turbulence. The major contribution of this work can be summarized as follows:
\begin{itemize}
\item We established the channel model and derived the BER of CD-Kennedy receiver under a $1\times N$ configuration with EGC method in turbulent channels.
\item We proposed the first post-selection based CV-QKD protocol with CD-Kennedy receiver for turbulent channels and derived the SKR expression.
\item We proposed an equivalent transmittance method to simplify the calculation of both the BER and SKR for the CD-Kennedy receivers with EGC method.
\item We demonstrate for the first time that the SKR performance advantage of using CD-Kennedy receiver over homodyne receiver becomes larger as the average transmittance decreases. Besides, the SKR performance of CD-Kennedy receiver is much robust than that of homodyne receiver in turbulent channels.
\item We also find that BER and SKR performance advantage of CD-Kennedy receiver over homodyne receiver demonstrate opposite trends as the average transmittance increases, which indicates that two separate system settings should be employed for communication and key distribution purposes.
\end{itemize}

The rest of this paper is organized as follows: we first derive the BER of both the CD-Kennedy receiver and the homodyne receiver with EGC method for a $1\times N$ configuration under turbulent channels in Section \ref{CD-Kennedy_BER}; then we derive the SKR of a post-selection based CV-QKD protocol by using both the CD-Kennedy receiver and the homodyne receiver under turbulent channels in Section \ref{CV-QKD_with_CD-Kennedy}; some numerical results on both the BER and SKR performance are presented in Section \ref{Numerical_Results} and we conclude our work in Section \ref{Conclusion}.

\section{CD-Kennedy receivers with equal-gain combining in turbulent channel}\label{CD-Kennedy_BER}

\begin{figure*}
\centering
\includegraphics[width=0.9\linewidth]{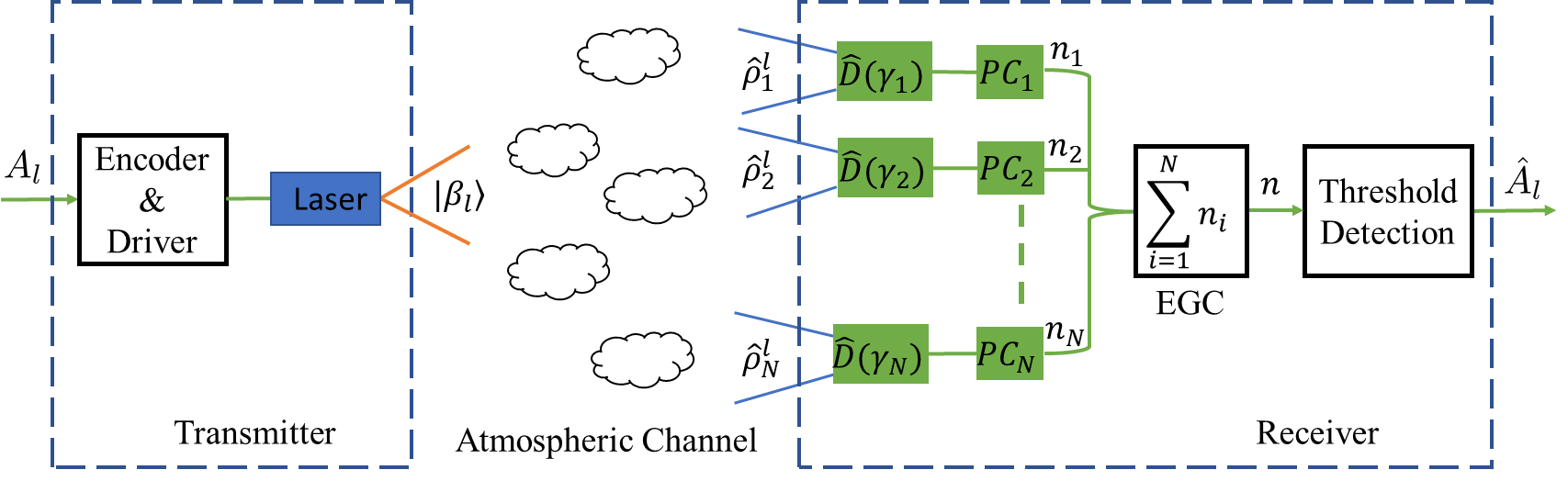}
\caption{Generalized Kennedy receivers with EGC in turbulent channel}
\label{system_set}
\end{figure*}

We consider a BPSK modulated communication system with $N$ CD-Kennedy receivers in this paper, as shown in Fig. \ref{system_set}. The transmitted bit $A_l \in \{0,1\}$ of the $l$th bit is encoded by a coherent state $\ket{\beta_l}$ with $\beta_l=-\beta$ for bit $A_l=0$ and $\beta_l=\beta$ for bit $A_l=1$. Without loss of generality, we adopt $\beta$ as real number. After passing through a turbulent channel, the transmitted state is received by $N$ receiving branches and the density operator of the signal arriving at the $i$th receiving branch is denoted by $\hat{\rho}^l_{i}$ with $i=1,2,\cdots,N$. We use $N$ CD-Kennedy receivers, whose displacement value $\gamma_i$ is dynamically conditioned on the average turbulent strength of the $i$th branch, to detect the received signals and the output numbers of photons at the $i$th photon counter ($PC_i$) is denoted by $n_i$. Then an EGC method is used to combine the output number of photons and the combined signal $n$ is decided by a threshold detection to recover the transmitted bit $\hat{A}_l$.

\subsection{Coherent states and $P$-representation}
The quantum state of a laser signal can be expressed in a coherent state $\ket{\alpha}$ with $\alpha\in \mathbb{C}$ and $\mathbb{C}$ is the field of complex number. By using the Fock basis of the Hilbert space, which consists of all number states (or Fock states) ${\{\ket{n},n=0, 1, 2, ...\}}$, we can expand the coherent $\ket{\alpha}$ as \cite{glauber1963quantum}
\begin{equation}
\label{Equa:CoherentState}
{\ket{\alpha}} = \sum_{n=0}^{\infty}e^{-\frac{1}{2}{|\alpha|}^2}\frac{{\alpha}^n}{\sqrt{n!}}\ket{n},
\end{equation}
\noindent where $|\alpha|^2$ is the average number of photons contained in the coherent state $\ket{\alpha}$.

According to Glauber's theory, all the coherent states ${\{\ket{\alpha}, \alpha \in \mathbb{C}\}}$ form an overcomplete basis of Hilbert space; and therefore, any quantum state with a density operator ${\hat{\rho}}$ in this Hilbert space could be decomposed by coherent states as \cite{glauber1963quantum}
\begin{equation}
\label{Equa:DensityfromP}
\hat{\rho} = \int_{\alpha}P(\alpha)\ket{\alpha}\bra{\alpha}\mathrm{d}^2\alpha,
\end{equation}
\noindent where $\mathrm{d}^2\alpha=\mathrm{d}Re(\alpha)\mathrm{d}Im(\alpha)$ and ${P(\alpha)}$ is the $P$-function of the density operator $\hat{\rho}$. This representation of density operator is called the $P$-representation \cite{glauber1963coherent}. Specifically, the $P$-function of a coherent state $\ket{\beta_l}$ can be expressed as a Dirac delta function
\begin{equation}
\delta^2(\alpha-\beta_l)\triangleq \delta\left(Re(\alpha)-Re(\beta_l)\right)\delta\left(Im(\alpha)-Im(\beta_l)\right).
\end{equation}

Considering the configuration in Fig. \ref{system_set}, the transmitted signal $\ket{\beta_l}$ for the $l$th bit are directed to $N$ separated receivers. Then the input state of the turbulent channel can be regarded as an $N$-mode coherent state with equal complex value $\alpha_1=\alpha_2=\cdots=\alpha_N\triangleq \frac{\beta_l}{\sqrt{N}}$. Then the $P$-function of the input state of the turbulent channel can be expressed by
\begin{equation}\label{Equa:InputStates}
P_{in}(\alpha_{1}, ..., \alpha_{N})=\prod_{i=1}^N \delta^2\left(\alpha_i-\frac{\beta_l}{\sqrt{N}}\right).
\end{equation}

\subsection{Relation between input and output $P$-functions in turbulent channel}
The relation between the input and output $P$-functions of turbulent channel for a $1\times 1$ configuration was first derived by Semenov \cite{semenov2009quantum,yuan2020closed}, and was later extend to a $1\times N$ configuration by Yuan \cite{yuan2020free} as
\begin{equation}\label{Equa:TurbulenceChannel}
P_{out}(\alpha_{1}, ..., \alpha_{N})=\int_{\bm{\eta}}\frac{p(\bm{\eta})}{\eta_{1}...\eta_{N}}P_{in}(\frac{\alpha_{1}}{\sqrt{\eta_{1}}}, ..., \frac{\alpha_{N}}{\sqrt{\eta_{N}}})\mathrm{d}\bm{\eta},
\end{equation}
\noindent where $\eta_i$ is the transmittance of the $i$th branch; $p(\bm{\eta})$ is the joint probability density function (PDF) of all transmittance $\bm{\eta}\triangleq [\eta_1,\eta_2,\cdots,\eta_N]^\text{T}$; $P_{in}(\alpha_1,\alpha_2,\cdots,\alpha_N)$ is the input $P$-function and $P_{out}(\alpha_1,\alpha_2,\cdots,\alpha_N)$ is the output $P$-function of the turbulent channel.

We adopt a log-normal distributed turbulent channel in this paper, where the PDF of the transmittance $\bm{\eta}$ satisfies the following joint log-normal PDF:
\begin{equation}
\label{Equa:LNdistribution}
\begin{aligned}
p(\bm{\eta})&=\frac{\exp\left(-\frac{1}{2}(\ln \bm{\eta}-\bm{\mu})^\text{T}\Sigma^{-1}(\ln \bm{\eta}-\bm{\mu})\right)}{\prod_{i=1}^{N}\eta_i\sqrt{(2\pi)^N \det(\Sigma)}},
\end{aligned}
\end{equation}
\noindent where $\bm{\mu}=[\mu_1,\mu_1,\cdots,\mu_N]^\text{T}$
 is the expectation of $\ln \bm{\eta}$; $\Sigma$ is the covariance matrix of $\ln \bm{\eta}$. Without loss of generality, we set $\mu_1=\mu_2=\cdots=\mu_N \triangleq \mu_0$ and assume that the correlation between arbitrary two branches are the same. Then the covariance matrix $\Sigma$ can be expressed as
 \begin{equation}
\Sigma=\sigma_0^2\left[
\begin{matrix}
 1      & \rho      & \cdots & \rho      \\
 \rho       & 1      & \cdots & \rho       \\
 \vdots & \vdots & \ddots & \vdots \\
 \rho       & \rho       & \cdots & 1      \\
\end{matrix}
\right],
\end{equation}
 \noindent where $\sigma_0^2$ is the variance of $\eta_i$ for $i=1,2,\cdots,N$ and $\rho$ is the correlation coefficient between $\ln(\eta_i)$ and $\ln(\eta_j)$ for $i\neq j$, i.e.,
\begin{equation}\label{rho}
\rho\triangleq \frac{\text{E}[\ln(\eta_i)\ln(\eta_j)]-\text{E}[\ln(\eta_i)]\text{E}[\ln(\eta_j)]}{\sqrt{\text{Var}[\ln(\eta_{i})]\text{Var}[\ln(\eta_{j})]}}.
\end{equation}
Therefore, the average transmittances of all branches are the same with each other, i.e., $\text{E}[\eta_1]=\text{E}[\eta_2]=\cdots=\text{E}[\eta_N]\triangleq \eta_0$, where $\eta_0$ can be obtained as
 \begin{equation}
 \eta_0=\exp\left(\mu_0+0.5\sigma_0^2\right).
 \end{equation}

Substituting \eqref{Equa:InputStates} into \eqref{Equa:TurbulenceChannel}, we can obtain the output $P$-function of the turbulent channel as \cite{yuan2020free}
\begin{equation}\label{Equa:PforTurbulence}
\begin{aligned}
P_{out}(\alpha_{1}, ..., \alpha_{N})=&(2N)^N \prod_{i=1}^N\left[\frac{\alpha_i}{\beta_l}\delta(Im(\alpha_i)Re(\beta_l)-Re(\alpha_i)Im(\beta_l))\right]\\
& \quad \times p_{tur}\left(N|\frac{\alpha_1}{\beta_l}|^2,N|\frac{\alpha_2}{\beta_l}|^2,\cdots, N|\frac{\alpha_N}{\beta_l}|^2\right).
\end{aligned}
\end{equation}

Consider the $i$th branch, the $P$-function can be obtained as
\begin{equation}\label{Equa:PforTurbulence_ith}
\begin{aligned}
P_{out}(\alpha_{i})&=\int_{\alpha_j\neq\alpha_i}P_{out}(\alpha_{1}, ..., \alpha_{N})\prod_{j\neq i}^N\mathrm{d}^2\alpha_j\\
&=\int_{\eta_i}\frac{p_{tur}(\eta_i)}{\eta_i}\delta^2\left(\frac{\alpha_i}{\sqrt{\eta_i}}-\frac{\beta_l}{\sqrt{N}}\right)\mathrm{d}\eta_i,
\end{aligned}
\end{equation}
\noindent where $p_{tur}(\eta_i)$ is the marginal PDF of $\eta_i$ obtained from $p_{tur}(\eta_i)=\int_{\eta_j\neq \eta_i}p_{tur}(\bm{\eta})\prod_{j\neq i}^N\mathrm{d}\eta_j$.

\subsection{CD-Kennedy receivers with EGC in turbulent channels}
\subsubsection{CD-Kennedy receivers with EGC}
We adopt the CD-Kennedy receiver structure proposed in \cite{yuan2020free} to mitigate the influence of turbulence, where the displacement value $\gamma_i$ of the $i$th branch is dynamically conditioned on the transmittance $\eta_i$ and the latter is estimated by pilot bits under slow-varying turbulent channels \cite{yuan2020free,zhu2002free}. Under this context, for a given time period, the transmittance on the $i$th branch can be regarded as a fixing value $\eta_i$. Then the $P$-function of the $i$th branch given transmittance $\eta_i$ and transmitted signal $\ket{\beta_l}$ can be obtained as
\begin{equation}\label{Equa:PforTurbulence_ith_given_eta_i}
\begin{aligned}
P_{out}(\alpha_{i}|\eta_i,\beta_l)&=\frac{1}{\eta_i}\delta^2\left(\frac{\alpha_i}{\sqrt{\eta_i}}-\frac{\beta_l}{\sqrt{N}}\right)\\
&=\delta^2\left(\alpha_i-\frac{\sqrt{\eta_i}\beta_l}{\sqrt{N}}\right),
\end{aligned}
\end{equation}
\noindent which corresponds to a coherent state $\ket{\frac{\sqrt{\eta_i}\beta_l}{\sqrt{N}}}$.

Then according to principle of generalized Kennedy receiver, we can use a displacement operator $\hat{D}(\gamma_i)$ with $\gamma_i=\frac{\sqrt{\eta_i}\beta}{\sqrt{N}}$ to displace the input state $\ket{\frac{\sqrt{\eta_i}\beta_l}{\sqrt{N}}}$ and use a photon-number resolving detector to detect the number of photons of the $i$th displaced state.
\begin{figure}
\centering
\includegraphics[width=0.5\linewidth]{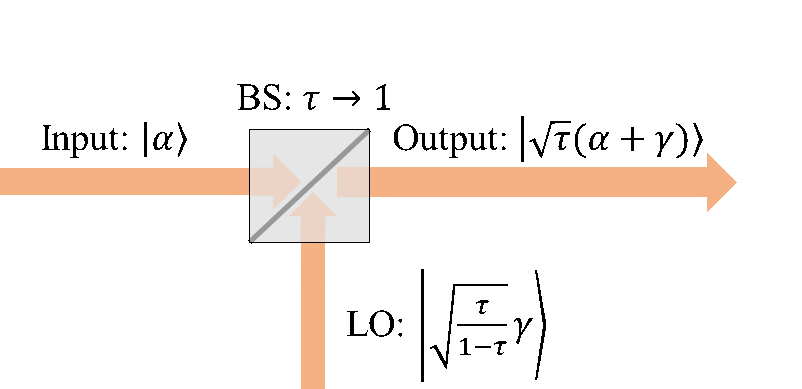}
\caption{Displacement operator in practical implementation}
\label{Displacement_operator}
\end{figure}

The displacement operation $\hat{D}(\gamma)$ can displace any coherent state $\ket{\alpha}$ to another coherent state $\hat{D}(\gamma)\ket{\alpha}=\ket{\alpha+\gamma}$. In practical implementation, the displacement operator can be achieved by combining the input coherent state with an local oscillator (LO) using a high transmittance beamsplitter (BS), as shown in Fig. \ref{Displacement_operator}. By using a BS with transmittance $\tau\to 1$ to combine the input coherent state $\ket{\alpha_{in}}=\ket{\alpha}$ and an LO state $\ket{\alpha_{LO}}=\ket{\sqrt{\frac{\tau}{1-\tau}}\gamma}$, we can obtain the output coherent state $\ket{\alpha_{out}}$ according to the property of the BS as
\begin{equation}
\begin{aligned}
\ket{\alpha_{out}}&=\ket{\sqrt{\tau}\alpha_{in}+\sqrt{1-\tau}\alpha_{LO}}\\
&=\ket{\sqrt{\tau}(\alpha+\gamma)},
\end{aligned}
\end{equation}
\noindent which will become the coherent state $\ket{\alpha+\gamma}$ as $\tau\to 1$.

Under this context, the detection probability of $n_i$ photons at the $i$th branch given $\eta_i$ and transmitted signal $\ket{\beta_l}$ can be obtained as a Poisson distribution:
\begin{equation}\label{Equa:photon_distribution_given_eta_i}
p(n_i|\eta_i,\beta_l)=\frac{\exp\left(-\left|\frac{\sqrt{\eta_i}(\beta_l+\beta)}{\sqrt{N}} \right|^2\right)}{n_i!}\left|\frac{\sqrt{\eta_i}(\beta_l+\beta)}{\sqrt{N}} \right|^{2n_i}.
\end{equation}

The output photon numbers are combined using the EGC combining with total output photon numbers given by
\begin{equation}
n=\sum_{i=1}^N n_i.
\end{equation}

Because $n_i$ satisfies Poisson distribution and the detection on each branches are independent with each other, the total output photon numbers $n$ given transmittance $\bm{\eta}$ and transmitted signal $\ket{\beta_l}$ also satisfies a Poisson distribution with probability density given by
\begin{equation}\label{Equa:photon_distribution_given_eta}
p(n|\bm{\eta},\beta_l)=\frac{\exp\left({-\sum_{i=1}^N\left|\frac{\sqrt{\eta_i}(\beta_l+\beta)}{\sqrt{N}} \right|^2}\right)}{n!}\left(\sum_{i=1}^N\left|\frac{\sqrt{\eta_i}(\beta_l+\beta)}{\sqrt{N}} \right|^{2}\right)^n.
\end{equation}

A maximum a posteriori (MAP) decision rule is used to decide the received bit as
\begin{equation}
\hat{A}_l=
\begin{cases}
0, \quad p_0 p(n|\bm{\eta},-\beta)\geq p_1 p(n|\bm{\eta},\beta),        \\
1, \quad p_0 p(n|\bm{\eta},-\beta)< p_1 p(n|\bm{\eta},\beta),
\end{cases}
\end{equation}
\noindent where $p_0$ and $p_1$ are the prior probabilities for transmitting bit 0 and 1, respectively. We consider equal prior probabilities in this work; then the above MAP decision rule will reduced to the following threshold detection as
\begin{equation}\label{decision_rule_CD_Kennedy}
\hat{A}_l=
\begin{cases}
0, \quad n=0,        \\
1, \quad n>0.
\end{cases}
\end{equation}

From \eqref{decision_rule_CD_Kennedy} we can see that the detection can be achieved by simple on/off photodetectors instead of expensive photon counters. Then the error probability of the CD-Kennedy receiver given transmittance $\bm{\eta}$ can be obtained as
\begin{equation}
P_e(\bm{\eta})=\frac{1}{2}\exp\left({-\frac{4\beta^2}{N}\sum_{i=1}^N \eta_i}\right).
\end{equation}

Finally, the unconditional error probability over the turbulent channel can be expressed as
\begin{equation}\label{P_e_CD_Kenndy}
P_e=\frac{1}{2}\int_{\bm{\eta}} p(\bm{\eta})\exp\left({-\frac{4\beta^2}{N}\sum_{i=1}^N \eta_i}\right)\mathrm{d}\bm{\eta},
\end{equation}
\noindent where $\mathrm{d}\bm{\eta}\triangleq \mathrm{d}\eta_1\mathrm{d}\eta_2\cdots\mathrm{d}\eta_N$.

It is challenging to calculate \eqref{P_e_CD_Kenndy} because it contains a $N$-tuple integral. However, by observing the expression of $P_e$ in \eqref{P_e_CD_Kenndy}, we can see that $\eta_i$ appears as an integrated term $\sum_{i=1}^N\eta_i$. Therefore, we can define an equivalent transmittance variable $\eta_{eq}\triangleq \sum_{i=1}^N\eta_i$ and rewrite \eqref{P_e_CD_Kenndy} as
\begin{equation}
P_e=\frac{1}{2}\int_{\eta_{eq}} p(\eta_{eq})\exp\left({-\frac{4\beta^2}{N}\eta_{eq}}\right)\mathrm{d}\eta_{eq},
\end{equation}
\noindent where $p(\eta_{eq})$ is the PDF of $\eta_{eq}$.

Since $\eta_{eq}$ is a summation of $N$ correlated log-normal random variables, we can approximate $\eta_{eq}$ as another log-normal random variable by using the Fenton-Wilkinson method \cite{cobb2012approximating}. Specifically, $\eta_{eq}$ can be approximated as a log-normal variable, i.e., $\eta_{eq}\sim LN(\mu_{eq},\sigma^2_{eq})$, where the parameters $\mu_{eq}$ and $\sigma^2_{eq}$ subject to the following two constrains:
\begin{equation}\label{two_constrains}
\begin{cases}
\int_{\eta_{eq}} \eta_{eq} p(\eta_{eq}) \mathrm{d} \eta_{eq}= \int_{\bm{\eta}}  (\sum_{i=1}^N\eta_i) p(\bm{\eta}) \mathrm{d} \bm{\eta}\\
\int_{\eta_{eq}} \eta_{eq}^2 p(\eta_{eq}) \mathrm{d} \eta_{eq}= \int_{\bm{\eta}} (\sum_{i=1}^N\eta_i)^2 p(\bm{\eta})  \mathrm{d} \bm{\eta}.
\end{cases}
\end{equation}

After some algebra (see \ref{Append_A}), we can obtain $\mu_{eq}$ and $\sigma^2_{eq}$ as
\begin{equation}\label{mu_eq_and_sigma_eq_2}
\begin{cases}
\mu_{eq}= 1.5\ln N+\mu_0-0.5\ln \left(1+(N-1)\exp\left((\rho-1)\sigma_0^2\right)\right),   \\
\sigma^2_{eq}=-\ln N+\sigma_0^2+\ln \left(1+(N-1)\exp\left((\rho-1)\sigma_0^2\right)\right).
\end{cases}
\end{equation}

\subsubsection{Homodyne receivers with EGC}
We use the homodyne receivers as the comparison. Then the output $x_i$ of the $i$th branch given transmittance $\eta_i$ and transmitted signal $\ket{\beta_l}$ can be equivalently modeled as a Gaussian distributed random variable with PDF given by
\begin{equation}
p(x_i|\eta_i,\beta_l)=\sqrt{\frac{2}{\pi}}\exp\left({-2\left(x_i-\frac{\sqrt{\eta_i}\beta_l}{\sqrt{N}}\right)^2}\right).
\end{equation}

After EGC combining, the total output $x=\sum_{i=1}^N x_i$ given transmittance $\bm{\eta}$ and transmitted signal $\ket{\beta_l}$ also satisfies a Gaussian distribution with PDF given by
\begin{equation}\label{p_x_homodyne}
p(x|\bm{\eta},\beta_l)=\sqrt{\frac{2}{N \pi}}\exp\left({-\frac{2}{N}\left(x-\sum_{i=1}^N\frac{\sqrt{\eta_i}\beta_l}{\sqrt{N}}\right)^2}\right).
\end{equation}

Similar to the CD-Kennedy receiver, when an MAP decision rule is adopted, the decision is equivalent to the following threshold detection as
\begin{equation}
\hat{A}_l=
\begin{cases}
0, \quad x\leq 0,        \\
1, \quad x>0.
\end{cases}
\end{equation}

Similarly, the unconditional error probability of homodyne receiver over turbulent channel can be obtained as
\begin{equation}
P_e^{hd}=\int_{\bm{\eta}} p(\bm{\eta}) Q\left(\sqrt{\frac{4\beta^2}{N^2}\left(\sum_{i=1}^N\sqrt{\eta_i}\right)^2}\right)\mathrm{d}\bm{\eta},
\end{equation}
\noindent where $Q(x)$ is the Q-function defined as $Q(x)\triangleq \frac{1}{\sqrt{2\pi}}\int_0^x \exp\left({-0.5 t^2}\right)\mathrm{d} t$.

Similarly, by observing the expression of $S_{kr}^{hd}(\beta)$ in \eqref{S_kr_beta_homodyne_1}, we can see that $\eta_i$ always appears as an integrated term $\sum_{i=1}^N\sqrt{\eta_i}$. Therefore, we can define an equivalent transmittance variable $\eta_{eq}^{hd}\triangleq \sum_{i=1}^N\sqrt{\eta_i}$ and rewrite $P_e^{hd}$ as
\begin{equation}\label{P_e_homodyne_2}
P_e^{hd}=\int_{\eta_{eq}^{hd}} p(\eta_{eq}^{hd}) Q\left(\sqrt{\frac{4\beta^2}{N^2}\left(\eta_{eq}^{hd}\right)^2}\right)\mathrm{d}\eta_{eq}^{hd},
\end{equation}
\noindent where $p(\eta_{eq}^{hd})$ is the PDF of $\eta_{eq}^{hd}$.

Since $\eta_{eq}^{hd}$ is a summation of $N$ correlated log-normal random variables $\{\sqrt{\eta_1},\sqrt{\eta_2},\cdots,\sqrt{\eta_N}\}$, we can also approximate $\eta_{eq}^{hd}$ as another log-normal random variable by using the Fenton-Wilkinson method. Similar to the derivation for the $\eta_{eq}$, we can obtain the parameters $\mu_{eq,hd}$ and $\sigma^2_{eq,hd}$ as (see \ref{Append_B})
\begin{equation}\label{mu_eq_and_sigma_eq_2_homodyne}
\begin{cases}
\mu_{eq,hd}= 1.5\ln N+0.5\mu_0-0.5\ln \left(1+(N-1)\exp\left(0.25(\rho-1)\sigma_0^2\right)\right),   \\
\sigma^2_{eq,hd}=-\ln N+0.25\sigma_0^2+\ln \left(1+(N-1)\exp\left(0.25(\rho-1)\sigma_0^2\right)\right).
\end{cases}
\end{equation}

\section{Post-selection based CV-QKD with CD-Kennedy receiver in turbulent channel}\label{CV-QKD_with_CD-Kennedy}
The coherent states are widely used in CV-QKD protols, where two legitimate users Alice and Bob use the coherent states to distribute secure keys over a public channel in the presence of potential eavesdropper Eve. In this section, we study the binary modulated CV-QKD protocol based on post-selection strategy by using CD-Kennedy receiver in turbulent channels.

\subsection{Binary modulated CV-QKD}
A typical binary modulated CV-QKD protocol consists of the following steps:
\begin{itemize}
\item Step 1: Alice randomly sends $\ket{-\beta}$ (bit 0) or $\ket{\beta}$ (bit 1) to Bob over a public channel;
\item Step 2: Bob measures the received state and decides the transmitted bit;
\item Step 3: Alice and Bob compare part of their bit strings and estimate the mutual information $I_{AB}$ between them and the mutual information $I_{AE}$ between Alice and Eve;
\item Step 4: Alice and Bob perform the information reconciliation using error correction methods;
\item Step 5: Alice and Bob perform the privacy amplification to extract at most $\Delta I = I_{AB}-I_{AE}$ bits of secret key.
\end{itemize}

The crucial step of the protocol is the estimate the mutual information $I_{AB}$ and $I_{AE}$. In the following we derive $I_{AB}$ and $I_{AE}$ over a turbulent channel when a CD-Kennedy receiver is used by Bob.

\subsection{Mutual information $I_{AB}$ between Alice and Bob}
\subsubsection{$I_{AB}$ with CD-Kennedy receivers}
We follow the procedure developed in \cite{silberhorn2002continuous} to derive the mutual information $I_{AB}$ for a transmitted signal amplitude $\beta$, where the channel is divided into many effective information channels characterized by the parameters $(n,\bm{\eta},\beta)$ with
\begin{equation}
I_{AB}(\beta)=\int_{\bm{\eta}}\sum_{n=0}^{\infty}p(\bm{\eta})p(n|\bm{\eta})I_{AB}(n,\bm{\eta},\beta)\mathrm{d}\bm{\eta},
\end{equation}
\noindent where $I_{AB}(n,\bm{\eta},\beta)$ is the effective information given the detected number of photons $n$, the measured transmittance $\bm{\eta}$, and the transmitted signal amplitude $\beta$; $p(n|\bm{\eta})$ is the probability of detecting $n$ photons given transmittance $\bm{\eta}$ at the receiver, which can be obtained by
\begin{equation}\label{p_n_eta}
\begin{aligned}
p(n|\bm{\eta})&=\frac{1}{2}p(n|\bm{\eta},-\beta)+\frac{1}{2}p(n|\bm{\eta},\beta)\\
&=\frac{1}{2}0^n+\frac{1}{2}\frac{\exp\left(-4\beta^2\sum_{i=1}^N\eta_i/N\right)}{n!}\left(4\beta^2\sum_{i=1}^N\eta_i/N\right)^n.
\end{aligned}
\end{equation}

For every detected number of photons $n$, Bob make a decision on the transmitted bit according to the decision rule given in \eqref{decision_rule_CD_Kennedy}; then the error rate of this decision can be obtained as
\begin{equation}
\begin{aligned}
p_e(n,\bm{\eta},\beta)&=
\begin{cases}
\frac{p(n|\bm{\eta},\beta)}{p(n|\bm{\eta},\beta)+p(n|\bm{\eta},-\beta)},\quad n=0,\\
\frac{p(n|\bm{\eta},-\beta)}{p(n|\bm{\eta},\beta)+p(n|\bm{\eta},-\beta)}, \quad n>0,
\end{cases}
\end{aligned}
\end{equation}
\noindent where $p(n|\bm{\eta},\beta_l)$ is given in \eqref{Equa:photon_distribution_given_eta}. Then the effective information $I_{AB}(n,\bm{\eta},\beta)$ can be obtained as
\begin{equation}\label{I_AB_n_eta_beta}
\begin{aligned}
I_{AB}(n,\bm{\eta},\beta)&=1-H(p_e(n,\bm{\eta},\beta))\\
&=1-H\left(\frac{0^n}{\exp\left(-4\beta^2\sum_{i=1}^N\eta_i/N\right)\left(4\beta^2\sum_{i=1}^N\eta_i/N\right)^n/n!+0^n}\right),
\end{aligned}
\end{equation}
\noindent where $H(p_e)\triangleq -p_e \log_2 p_e-(1-p_e)\log_2(1-p_e)$ is the entropy of the effective information channel.

\subsubsection{$I_{AB}$ with homodyne receivers}
For comparison, here we give the mutual information between Alice and Bob when a homodyne receiver is employed, i.e.,
\begin{equation}
I_{AB}^{hd}(\beta)=\int_{\bm{\eta}}\int_x p(\bm{\eta}) p(x|\bm{\eta})I_{AB}^{hd}(x,\bm{\eta},\beta)\mathrm{d}\bm{\eta}\mathrm{d} x,
\end{equation}
\noindent where $p(x|\bm{\eta})$ is the probability of output $x$ given transmittance $\bm{\eta}$ and $I_{AB}^{hd}(x,\bm{\eta},\beta)$ is the effective information given output $x$, transmittance $\bm{\eta}$, and signal amplitude $\beta$.

By using eq. \eqref{p_x_homodyne}, we can obtain $p(x|\bm{\eta})$ as
\begin{equation}
\begin{aligned}
p(x|\bm{\eta})&=\frac{1}{2}p(x|\bm{\eta},-\beta)+\frac{1}{2}p(x|\bm{\eta},\beta)\\
&=\frac{1}{2}\sqrt{\frac{2}{N \pi}}\left[\exp\left({-\frac{2}{N}\left(x+\frac{\beta}{\sqrt{N}}\sum_{i=1}^N\sqrt{\eta_i} \right)^2}\right)\right.\\
&\quad\quad\quad\quad\quad \left.+\exp\left({-\frac{2}{N}\left(x-\frac{\beta}{\sqrt{N}}\sum_{i=1}^N\sqrt{\eta_i} \right)^2}\right)\right].
\end{aligned}
\end{equation}

Similar to the CD-Kennedy receiver, the error probability when an output $x$ is measured given transmittance $\bm{\eta}$ and signal amplitude $\beta$ can be obtained as
\begin{equation}
\begin{aligned}
p_e(x,\bm{\eta},\beta)&=
\begin{cases}
\frac{p(x|\bm{\eta},\beta)}{p(n|\bm{\eta},\beta)+p(x|\bm{\eta},-\beta)},\quad x\leq 0,\\
\frac{p(x|\bm{\eta},-\beta)}{p(n|\bm{\eta},\beta)+p(x|\bm{\eta},-\beta)}, \quad x>0.
\end{cases}
\end{aligned}
\end{equation}

Then the effective information  $I_{AB}^{hd}(x,\bm{\eta},\beta)$ for homodyne receiver can be obtained as
\begin{equation}
\begin{aligned}
I_{AB}^{hd}(x,\bm{\eta},\beta)&=1-H(p_e(x,\bm{\eta},\beta))\\
&=1-H\left(\frac{1}{\exp\left(-8x\beta\sum_{i=1}^N\sqrt{\eta_i}/N^{3/2}\right)+1}\right).
\end{aligned}
\end{equation}

\subsection{Mutual information $I_{AE}$ between Alice and Eve}
The mutual information between Alice and Eve depends on the attack methods of Eve. For a channel with negligible excess noise, the best eavesdropping strategy of Eve is the passive beamsplitter attack \cite{heid2006efficiency,sych2010coherent}. In a turbulent channel, we further assume that Eve can split the beam near the transmitter and thus Eve can safely split $(1-\bar{\eta})$ quantity of the beam energy without being discovered, where $\bar{\eta}=N\eta_0$ is the average beam energy detected by the receiver. Then Eve has to discriminate two coherent states
\begin{equation}
\left\{\ket{-\sqrt{1-\bar{\eta}}\beta},\ket{\sqrt{1-\bar{\eta}}\beta}\right\}.
\end{equation}

For an individual attack, Eve decides each bit individually, then the minimum error rate of Eve is obtained by the Helstrom's theory as \cite{helstrom1969quantum,helstrom1970quantum}
\begin{equation}
p_e(\beta)=\frac{1}{2}\left(1-\sqrt{1-|f|^2}\right),
\end{equation}
\noindent where $f\triangleq \langle -\sqrt{1-\bar{\eta}}\beta | \sqrt{1-\bar{\eta}}\beta \rangle=e^{-2(1-N\eta_0)\beta^2}$.

Then the mutual information $I_{AE}$ under individual attack can be obtained by
\begin{equation}
\begin{aligned}
I_{AE}(\beta)&=1-H(p_e(\beta))\\
&=\frac{1}{2}(1-\sqrt{1-f^2})\log_2(1-\sqrt{1-f^2})\\
&\quad +\frac{1}{2}(1+\sqrt{1-f^2})\log_2(1+\sqrt{1-f^2}).
\end{aligned}
\end{equation}

For a collective attack, Eve can collect the splitted bits and make decision over all collected bits. Then the mutual information $I_{AE}$ is given by the Holevo bound as
\begin{equation}
\begin{aligned}
I_{AE}(\beta)&=1-\frac{1}{2}(1-f)\log_2(1-f)-\frac{1}{2}(1+f)\log_2(1+f).
\end{aligned}
\end{equation}

\subsection{Post-selection strategy in turbulent channel}
\subsubsection{Post-selection strategy for CD-Kennedy receivers}
Because Bob can only access to $\bar{\eta}$ quantity of the transmitted beam energy, as long as $\bar{\eta}<0.5$, Eve can always access more knowledge of the transmitted bits, which lead to the ``3dB loss limit" of the binary modulated CV-QKD protocol \cite{grosshans2002continuous}. To achieve an advantage over Eve, Bob can only save those bits with higher effective information $I_{AB}(n,\bm{\eta},\beta)$ than $I_{AE}$ and discard those bits with lower effective information. This is the so called post-selection strategy for beating the 3dB loss limit, which is first proposed in \cite{silberhorn2002continuous}.

For a turbulent channel, Bob can save those bits with $I_{AB}(n,\bm{\eta},\beta)\geq I_{AE}(\beta)$, which corresponds to a post-selection area $\mathbf{A_{ps}}$ defined as
\begin{equation}
\mathbf{A_{ps}}=\{(n,\bm{\eta})|I_{AB}(n,\bm{\eta},\beta)\geq I_{AE}(\beta)\}.
\end{equation}

Then the secret key rate for a given transmitted signal amplitude $\beta$ can be obtained as
\begin{equation}\label{S_kr_beta_1}
\begin{aligned}
S_{kr}(\beta)&=\int_{\mathbf{A_{ps}}} p(n|\bm{\eta}) p(\bm{\eta}) (I_{AB}(n,\bm{\eta},\beta)- I_{AE}(\beta)) \mathrm{d}\bm{\eta}.
\end{aligned}
\end{equation}

It is challenging to find the post-selection area ${\mathbf{A_{ps}}}$ directly because  ${\mathbf{A_{ps}}}$ is a $(N+1)$-dimensional area. However, by observing the expression of $S_{kr}(\beta)$ in \eqref{S_kr_beta_1}, we can see that $\eta_i$ always appears as an integrated term $\sum_{i=1}^N\eta_i$. Therefore, similar to the calculation of the error probability, we can define an equivalent transmittance variable $\eta_{eq}\triangleq \sum_{i=1}^N\eta_i$ and rewrite $S_{kr}(\beta)$ as
\begin{equation}\label{S_kr_beta_2}
\begin{aligned}
S_{kr}(\beta)&=\int_{\mathbf{A_{ps,eq}}} p(\eta_{eq}) p(n|\eta_{eq}) (I_{AB}(n,\eta_{eq},\beta)- I_{AE}(\beta)) \mathrm{d}\eta_{eq},
\end{aligned}
\end{equation}
\noindent where $p(n|\eta_{eq})$ and $I_{AB}(n,\eta_{eq},\beta)$ are defined as
\begin{equation}\label{p_n_eta_eq_and_I_AB_n_eta_eq}
\begin{cases}
p(n|\eta_{eq})\triangleq \frac{1}{2}\left[0^n+\exp\left(-\frac{4\beta^2}{N}\eta_{eq}\right)\left(\frac{4\beta^2}{N}\eta_{eq}\right)^n/n!\right]\\ I_{AB}(n,\eta_{eq},\beta)\triangleq 1-H\left(\frac{0^n}{{\exp\left(-\frac{4\beta^2}{N}\eta_{eq}\right)\left(\frac{4\beta^2}{N}\eta_{eq}\right)^n}/{n!}+0^n}\right).
\end{cases}
\end{equation}

Then the post-selection area ${\mathbf{A_{ps}}}$ becomes an equivalent post-selection area ${\mathbf{A_{ps,eq}}}$ with only two dimensions $n$ and $\eta_{eq}$:
\begin{equation}
\mathbf{A_{ps,eq}}=\left\{(n,\eta_{eq})|I_{AB}(n,\eta_{eq},\beta)\geq I_{AE}(\beta)\right\}.
\end{equation}

Substituting eqs. \eqref{p_n_eta_eq_and_I_AB_n_eta_eq} and \eqref{mu_eq_and_sigma_eq_2} into \eqref{S_kr_beta_2}, we can obtain the secret key rate as
\begin{equation}\label{S_kr_beta_3}
\begin{aligned}
S_{kr}(\beta)&=\frac{1}{2}\int_{\mathbf{A_{ps,eq}}}p(\eta_{eq}) \left[0^n+\frac{\exp\left(-\frac{4\beta^2}{N}\eta_{eq}\right)\left(\frac{4\beta^2}{N}\eta_{eq}\right)}{n!}^n\right]\\
&\quad \times \left[ 1-H\left(\frac{0^n}{\frac{\exp\left(-\frac{4\beta^2}{N}\eta_{eq}\right)\left(\frac{4\beta^2}{N}\eta_{eq}\right)^n}{n!}+0^n}\right)- I_{AE}(\beta)\right] \mathrm{d}\eta_{eq}.
\end{aligned}
\end{equation}

\subsubsection{Post-selection strategy for homodyne receivers}
Here we present the post-selection strategy for homodyne receivers with EGC in turbulent channel. Similar to the CD-Kennedy receivers, Bob can save those bits with $I_{AB}^{hd}(x,\bm{\eta},\beta)\geq I_{AE}(\beta)$, which corresponds to a post-selection area $\mathbf{A_{ps}^{hd}}$ defined as
\begin{equation}
\mathbf{A_{ps}^{hd}}=\{(x,\bm{\eta})|I_{AB}^{hd}(x,\bm{\eta},\beta)\geq I_{AE}(\beta)\}.
\end{equation}

Then the secret key rate for a given transmitted signal amplitude $\beta$ can be obtained as
\begin{equation}\label{S_kr_beta_homodyne_1}
\begin{aligned}
S_{kr}^{hd}(\beta)&=\int_{\mathbf{A_{ps}^{hd}}} p(x|\bm{\eta}) p(\bm{\eta}) (I_{AB}^{hd}(x,\bm{\eta},\beta)- I_{AE}(\beta)) \mathrm{d}\bm{\eta}\mathrm{d}x.
\end{aligned}
\end{equation}

Similarly, by observing the expression of $S_{kr}^{hd}(\beta)$ in \eqref{S_kr_beta_homodyne_1}, we can see that $\eta_i$ always appears as an integrated term $\sum_{i=1}^N\sqrt{\eta_i}$. Therefore, we can define an equivalent transmittance variable $\eta_{eq}^{hd}\triangleq \sum_{i=1}^N\sqrt{\eta_i}$ and rewrite $S_{kr}^{hd}(\beta)$ as
\begin{equation}\label{S_kr_beta_homodyne_2}
\begin{aligned}
S_{kr}^{hd}(\beta)&=\int_{\mathbf{A_{ps,eq}^{hd}}} p(\eta_{eq}^{hd}) p(x|\eta_{eq}^{hd}) (I_{AB}^{hd}(x,\eta_{eq}^{hd},\beta)- I_{AE}(\beta)) \mathrm{d}\eta_{eq}^{hd}\mathrm{d}x.,
\end{aligned}
\end{equation}
\noindent where $p(x|\eta_{eq}^{hd}) $ and $I_{AB}^{hd}(x,\eta_{eq}^{hd},\beta)$ are defined as
\begin{equation}\label{p_x_eta_eq_and_I_AB_x_eta_eq}
\begin{cases}
p(x|\eta_{eq}^{hd})\triangleq \frac{1}{2}\sqrt{\frac{2}{N \pi}}\left[\exp\left({-\frac{2}{N}\left(x+\frac{\beta}{\sqrt{N}}\eta_{eq}^{hd} \right)^2}\right)\right.\\
\quad\quad\quad\quad\quad\quad\quad\left.+\exp\left({-\frac{2}{N}\left(x-\frac{\beta}{\sqrt{N}}\eta_{eq}^{hd} \right)^2}\right)\right].
\\ I_{AB}^{hd}(x,\eta_{eq}^{hd},\beta) \triangleq 1-H\left(\frac{1}{\exp\left(-8x\beta\eta_{eq}^{hd}/N^{3/2}\right)+1}\right);
\end{cases}
\end{equation}
\noindent and $p(\eta_{eq}^{hd})$ is the PDF of $\eta_{eq}^{hd}$. Then the post-selection area ${\mathbf{A_{ps}}}$ becomes an equivalent post-selection area ${\mathbf{A_{ps,eq}^{hd}}}$ with only two dimensions $x$ and $\eta_{eq}^{hd}$:
\begin{equation}
\mathbf{A_{ps,eq}^{hd}}=\left\{(x,\eta_{eq}^{hd})|I_{AB}^{hd}(x,\eta_{eq}^{hd},\beta)\geq I_{AE}(\beta)\right\}.
\end{equation}

Substituting eqs. \eqref{p_x_eta_eq_and_I_AB_x_eta_eq} and \eqref{mu_eq_and_sigma_eq_2_homodyne} into \eqref{S_kr_beta_homodyne_2}, we can obtain the secret key rate of homodyne receivers as
\begin{equation}\label{S_kr_beta_homodyn_3}
\begin{aligned}
&S_{kr}^{hd}(\beta)\\
&\quad=\frac{1}{2}\sqrt{\frac{2}{N \pi}}\int_{\mathbf{A_{ps,eq}^{hd}}} p(\eta_{eq}^{hd}) \left[ 1-H\left(\frac{1}{\exp\left(-8x\beta\eta_{eq}^{hd}/N^{3/2}\right)+1}\right)\right]\\
&\quad \quad \times \left[\exp\left({-\frac{2}{N}\left(x+\frac{\beta}{\sqrt{N}}\eta_{eq}^{hd} \right)^2}\right)+\exp\left({-\frac{2}{N}\left(x-\frac{\beta}{\sqrt{N}}\eta_{eq}^{hd} \right)^2}\right)\right]\mathrm{d}\eta_{eq}^{hd}\mathrm{d}x.
\end{aligned}
\end{equation}
\section{Numerical results}\label{Numerical_Results}

In this section we present some numerical results to explore both the BER and the SKR performance of CD-Kennedy receiver with EGC in turbulent channels. The homodyne receiver is employed as the comparison scheme. Unless otherwise specified, we set the average signal photons $|\beta|^2=2$, number of branches $N=4$, turbulent strength $\sigma_0^2=0.1$, and turbulent correlation coefficient $\rho=0$.

\begin{figure}
\begin{center}
\subfigure[]{\includegraphics[width=0.5\textwidth]{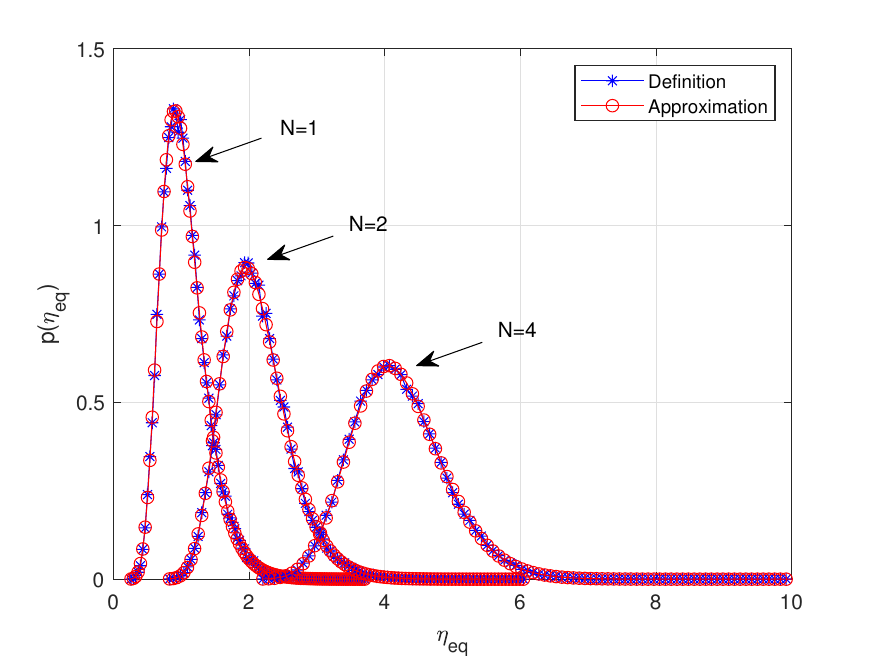}\label{Fig:eta_eq_approx}}
\subfigure[]{\includegraphics[width=0.5\textwidth]{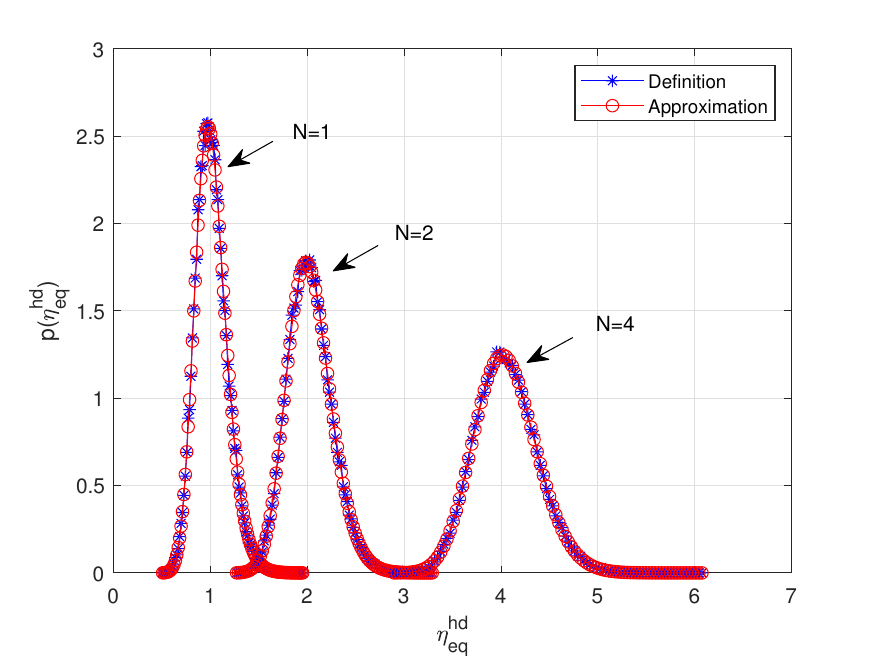}\label{Fig:eta_eq_hd_approx}}
\caption{PDF comparison between the definition and approximation of equivalent transmittance: (a) PDF of $\eta_{eq}$; (b) PDF of $\eta_{eq}^{hd}$}
\label{Fig:eta_eq_verify}
\end{center}
\end{figure}

We first verify the feasibility of using approximated equivalent transmittances $\eta_{eq}$ and $\eta_{eq}^{hd}$ by Fenton-Wilkinson method. Figs. \ref{Fig:eta_eq_approx} and \ref{Fig:eta_eq_hd_approx} present the PDF of $\eta_{eq}$ and $\eta_{eq}^{hd}$ under different number of branches with $N=1,2,4$, respectively. From Figs. \ref{Fig:eta_eq_approx} and \ref{Fig:eta_eq_hd_approx}, we can observe that the PDF approximations of both $\eta_{eq}$ and $\eta_{eq}^{hd}$ by using  Fenton-Wilkinson method can well match those of the definitions. Besides, the results in Figs. \ref{Fig:eta_eq_approx} and \ref{Fig:eta_eq_hd_approx} also verify the correctness of our derivation in \eqref{mu_eq_and_sigma_eq_2} and \eqref{mu_eq_and_sigma_eq_2_homodyne}.

\subsection{Bit-error rate comparison}

\begin{figure}
\centering
\includegraphics[width=0.5\textwidth]{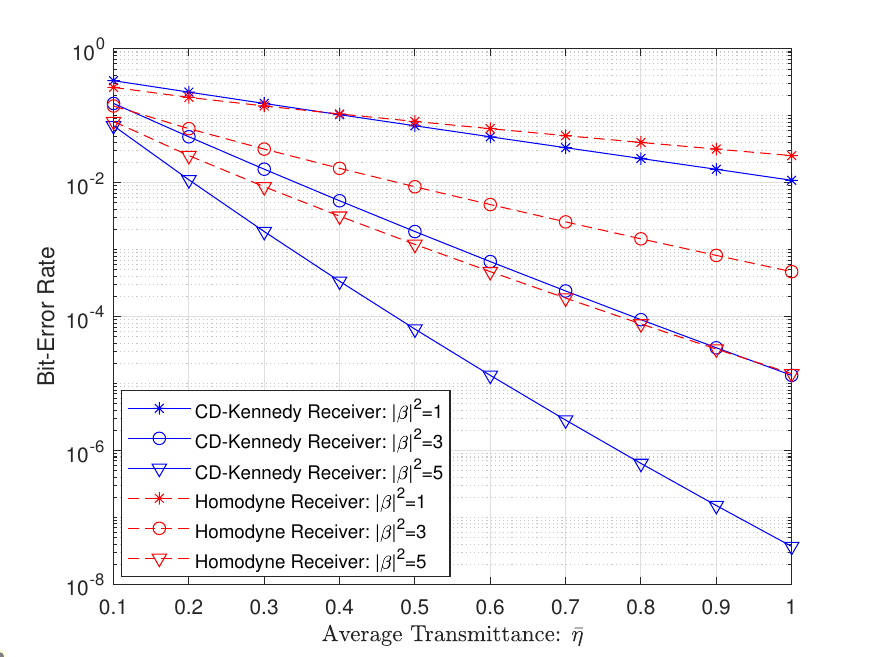}
\caption{BER comparison between CD-Kennedy receiver and homodyne receiver in turbulent channels under different transmitting signal strength $|\beta|^2$}
\label{Fig:BER_Different_beta_2}
\end{figure}

We then present the BER comparison between CD-Kennedy receiver and homodyne receiver in turbulent channels at different transmitting signal strength $|\beta|^2$ in Fig. \ref{Fig:BER_Different_beta_2}. From Fig. \ref{Fig:BER_Different_beta_2}, we can see that the BER performance of CD-Kennedy receiver with EGC is better than that of homodyne receiver in turbulent channels when the signal strength $|\beta|^2$ is large. Besides, for a given signal strength, e.g., when $|\beta|^2>1$, the performance advantage of CD-Kennedy receiver over homodyne receiver increases as average transmittance $\bar{\eta}$ increases.

\begin{figure}
\begin{center}
\subfigure[]{\includegraphics[width=0.5\textwidth]{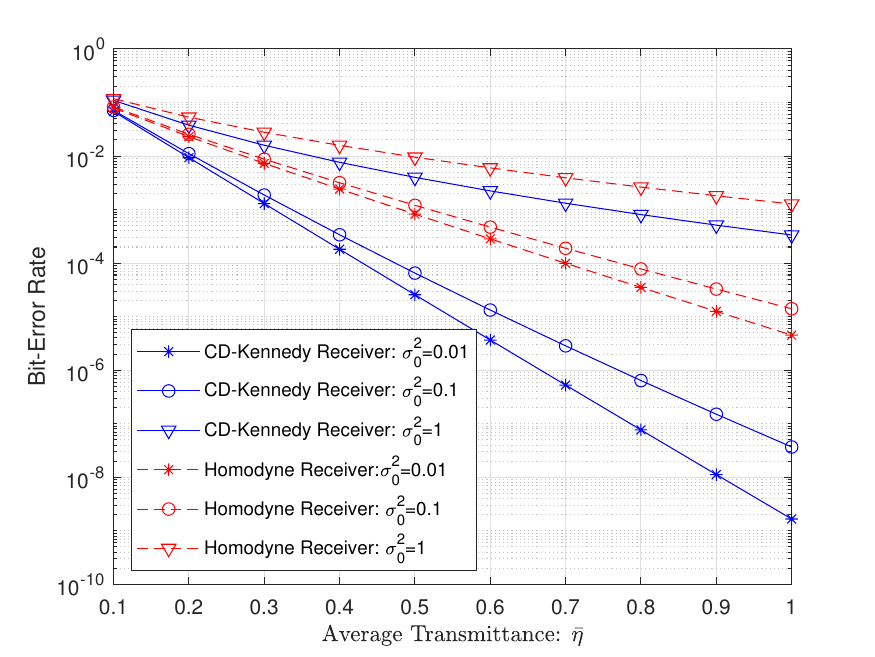}\label{Fig:BER_Different_Sigma_0_2}}
\subfigure[]{\includegraphics[width=0.5\textwidth]{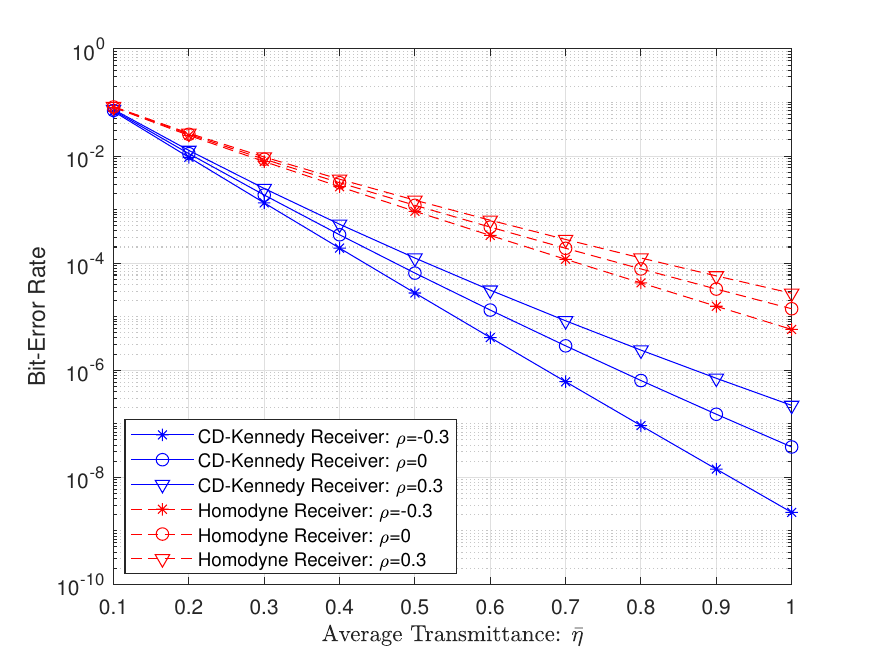}\label{Fig:BER_Different_rho}}
\caption{BER comparison between CD-Kennedy receiver and homodyne receiver in turbulent channels: (a) Different turbulent strength $\sigma_0^2$; (b) Different correlation coefficient $\rho$}
\label{Fig:BER_Compare_Different_Turbulence_Condition}
\end{center}
\end{figure}

Then we present the BER comparison between CD-Kennedy receiver and homodyne receiver in turbulent channels at different turbulent conditions, where Figs. \ref{Fig:BER_Different_Sigma_0_2} and \ref{Fig:BER_Different_rho} show the BER comparison under different turbulent strength $\sigma_0^2$ and different correlation coefficient $\rho$, respectively. From Figs. \ref{Fig:BER_Different_Sigma_0_2} and \ref{Fig:BER_Different_rho} we can see that the BER performance of CD-Kennedy receiver with EGC is better than that of homodyne receiver in turbulent channels when the average transmittance $\bar{\eta}$ is large. Besides, the advantage of CD-Kennedy receiver with EGC over homodyne receiver becomes large as the average transmittance $\bar{\eta}$ increases.

\begin{figure}
\centering
\includegraphics[width=0.5\textwidth]{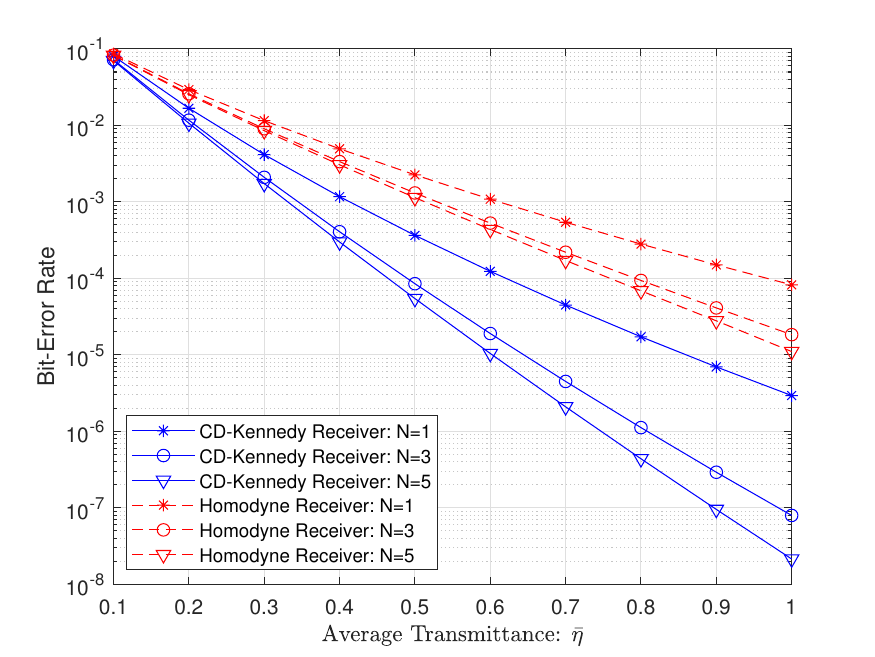}
\caption{BER comparison between CD-Kennedy receiver and homodyne receiver in turbulent channels under different number of receivers $N$}
\label{Fig:BER_Different_N}
\end{figure}

Then we present the BER comparison between CD-Kennedy receiver and homodyne receiver in turbulent channels at different number of receivers $N$ in Fig. \ref{Fig:BER_Different_N}. From Fig. \ref{Fig:BER_Different_N}, we can see that the BER performance of CD-Kennedy receiver is better than that of homodyne receiver. Besides, we can also see that the BER performance of both CD-Kennedy receiver and homodyne receiver improves as the number of branches increases, which indicates a receiving diversity gain is achieved.

\subsection{Secret key rate comparison}

\begin{figure}
\centering
\includegraphics[width=0.5\textwidth]{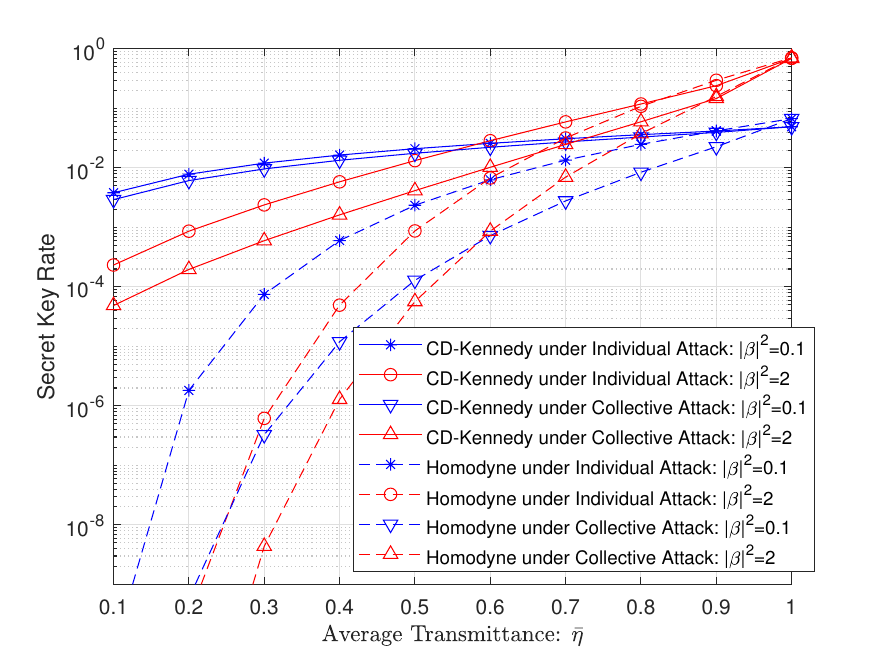}
\caption{SKR comparison between CD-Kennedy receiver and homodyne receiver in turbulent channels under different transmitting signal strength $|\beta|^2$}
\label{Fig:SKR_Different_beta_2}
\end{figure}

Then we present the SKR comparison between CD-Kennedy receiver and homodyne receiver in turbulent channels at different transmitting signal strength $|\beta|^2$ in Fig. \ref{Fig:SKR_Different_beta_2}. From Fig. \ref{Fig:SKR_Different_beta_2}, we can observe that for a given channel average transmittance, different transmitting signal strength $|\beta|^2$ can result in different SKR, which indicates there exits an optimum transmitting signal strength for a given average transmittance.

\begin{figure}
\begin{center}
\subfigure[]{\includegraphics[width=0.5\textwidth]{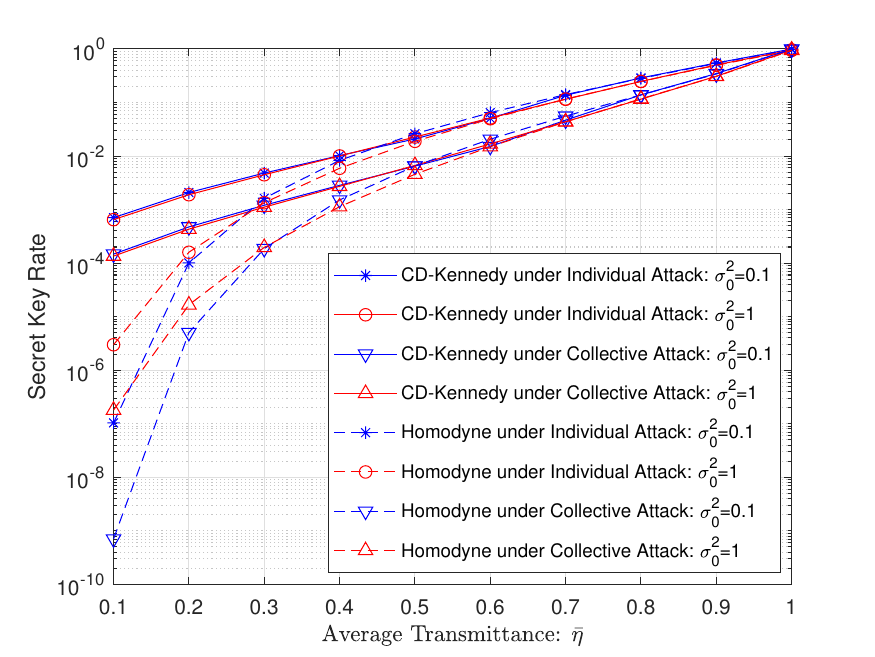}\label{Fig:SKR_Different_Sigma_0_2}}
\subfigure[]{\includegraphics[width=0.5\textwidth]{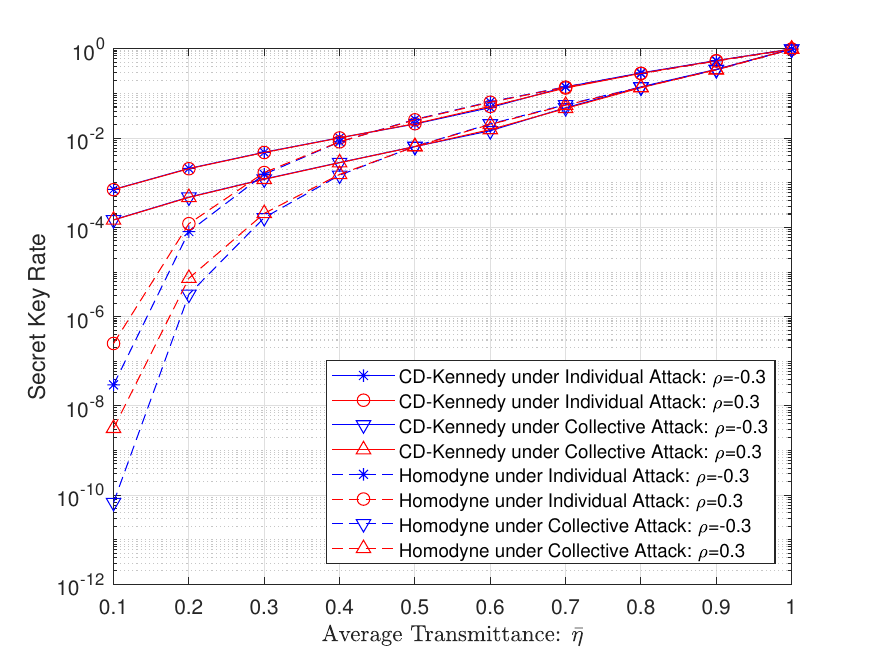}\label{Fig:SKR_Different_rho}}
\caption{SKR comparison between CD-Kennedy receiver and homodyne receiver in turbulent channels: (a) Different turbulent strength $\sigma_0^2$; (b) Different correlation coefficient $\rho$}
\label{Fig:SKR_Compare_Different_Turbulence_Condition}
\end{center}
\end{figure}

Then we present the SKR comparison between CD-Kennedy receiver and homodyne receiver in turbulent channels at different turbulent conditions in Fig. \ref{Fig:SKR_Compare_Different_Turbulence_Condition}, where Figs. \ref{Fig:SKR_Different_Sigma_0_2} and \ref{Fig:SKR_Different_rho} show the SKR comparison under different turbulent strength $\sigma_0^2$ and different correlation coefficient $\rho$, respectively. From Figs. \ref{Fig:SKR_Different_Sigma_0_2} and \ref{Fig:SKR_Different_rho}, we can see that the SKR performance of CD-Kennedy receiver is much robust than the homodyne receiver in turbulent channels. Besides, we can also see that, the SKR performance advantage of CD-Kennedy receiver over homodyne receiver increases as the average transmittance decreases, which demonstrate an opposite trend compared with the BER performance cases. This is because in a post-selection strategy, the more uncertainty of the channel, the more chance a better effective information can be achieved.

\begin{figure}
\centering
\includegraphics[width=0.5\textwidth]{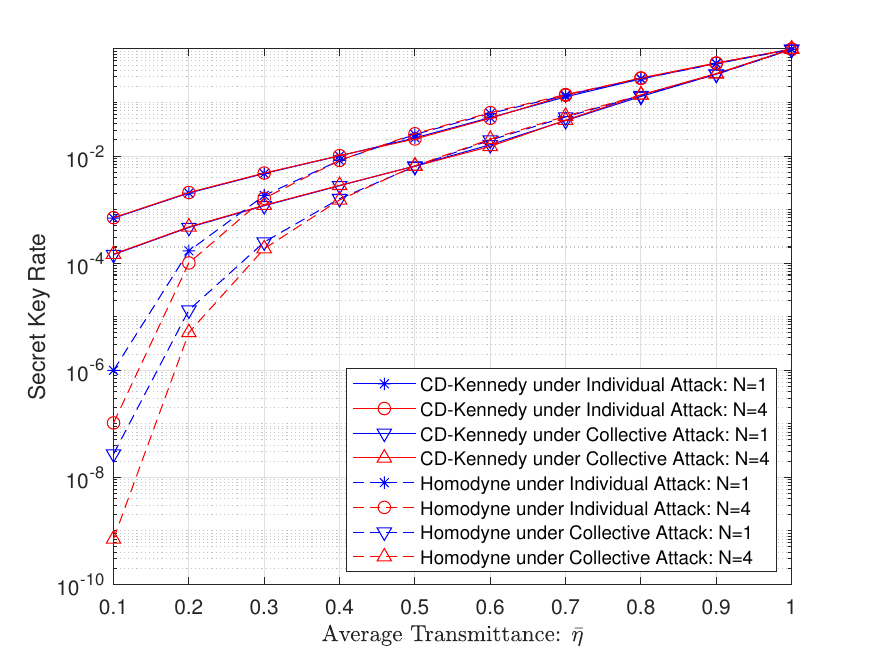}
\caption{SKR comparison between CD-Kennedy receiver and homodyne receiver in turbulent channels under different number of receivers $N$}
\label{Fig:SKR_Different_N}
\end{figure}

Then we present the SKR comparison between CD-Kennedy receiver and homodyne receiver in turbulent channels at different number of receivers $N$ in Fig. \ref{Fig:SKR_Different_N}. From \ref{Fig:SKR_Different_N} we can see that the SKR performance of post-selection based CD-QKD protocol with homodyne receiver deteriorates as the number of branches increases, which also demonstrate an opposite trend compared with the BER performance cases. This observation suggests that two separate system setting should be adopted for communication and key distribution purposes.

\begin{figure}
\begin{center}
\subfigure[]{\includegraphics[width=0.5\textwidth]{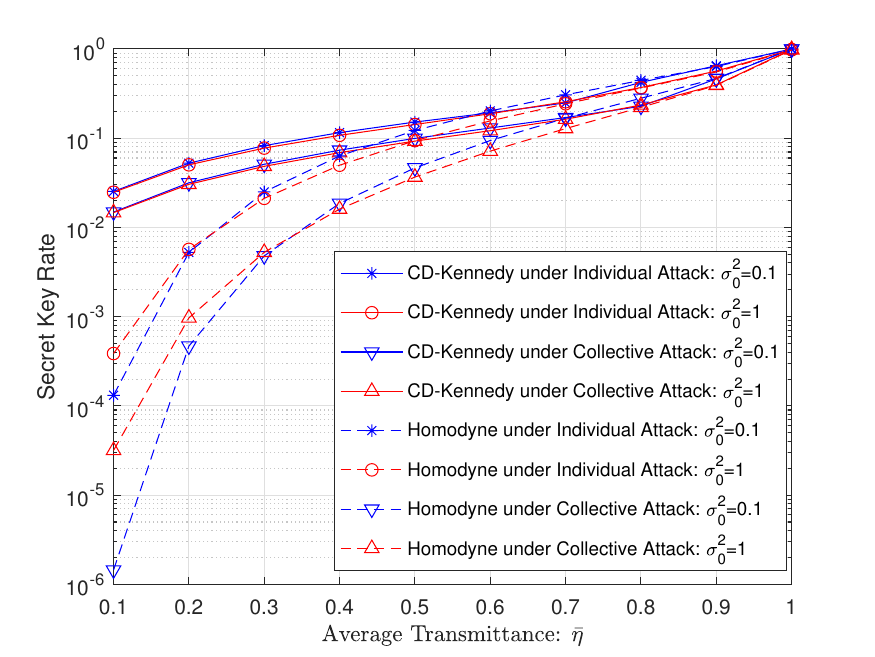}\label{Fig:SKR_Different_Sigma_0_2_optimal_beta}}
\subfigure[]{\includegraphics[width=0.5\textwidth]{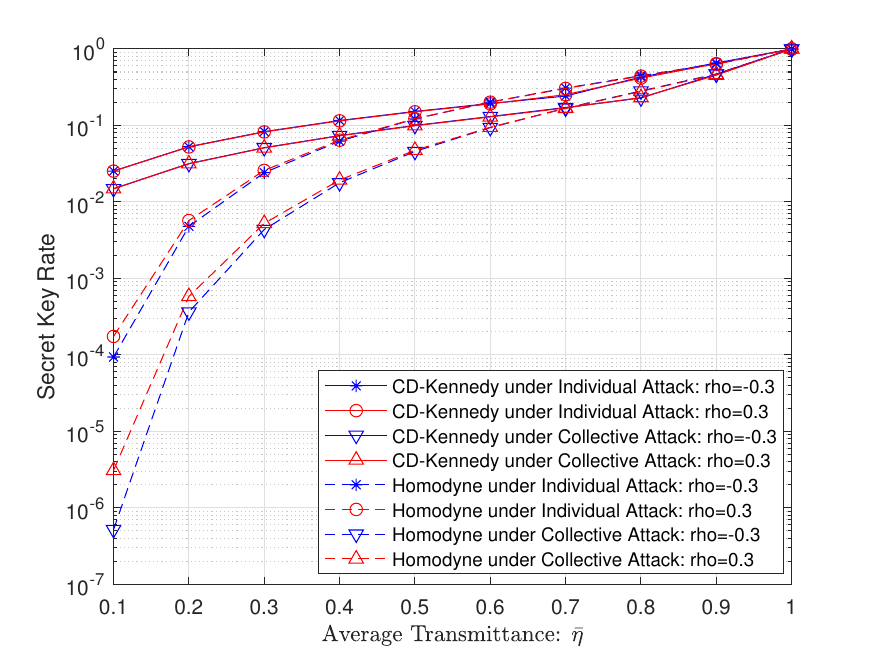}\label{Fig:SKR_Different_rho_optimal_beta}}
\subfigure[]{\includegraphics[width=0.5\textwidth]{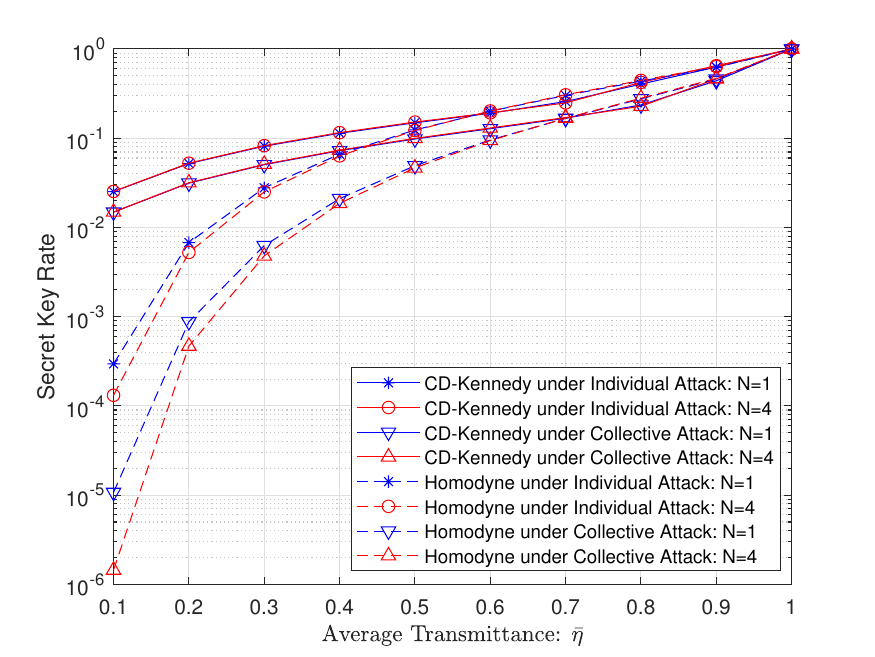}\label{Fig:SKR_Different_N_optimal_beta}}
\caption{SKR comparison between CD-Kennedy receiver and homodyne receiver with optimized transmitting signal strength $|\beta|^2$ in turbulent channels: (a) Different turbulent strength $\sigma_0^2$; (b) Different correlation coefficient $\rho$; (c) Different number of receivers $N$}
\label{Fig:Optimal_SKR_Compare}
\end{center}
\end{figure}

At last, we present the SKR comparison between CD-Kennedy receiver and homodyne receiver in turbulent channels with optimized transmitting signal strength at different turbulent strength $\sigma_0^2$, correlation coefficient $\rho$, and number of receivers $N$ in Figs. \ref{Fig:SKR_Different_Sigma_0_2_optimal_beta}, \ref{Fig:SKR_Different_rho_optimal_beta}, and \ref{Fig:SKR_Different_N_optimal_beta}, respectively. From Figs. \ref{Fig:SKR_Different_Sigma_0_2_optimal_beta} and \ref{Fig:SKR_Different_rho_optimal_beta}, we can see that the SKR performance with optimized signal strength for CD-Kennedy receiver is much robust than the homodyne receiver in turbulent channels. Besides, we can also see that, the SKR performance advantage of CD-Kennedy receiver over homodyne receiver increases as the average transmittance decreases. From \ref{Fig:SKR_Different_N_optimal_beta} we can see that the SKR performance with optimized signal strength for post-selection based CD-QKD protocol using homodyne receiver deteriorates as the number of branches increases.

\section{Conclusion}\label{Conclusion}

Endogenous security plays a crucial role in space-air-ground integrated network and a typical solution to endogenous security problems is the QKD. Compared with DV-QKD, CV-QKD enjoys higher SKR and better compatibility with current optical communication infrastructure. In this paper, we employed the CD-Kennedy receiver to enhance the detection performance of coherent states in turbulent channels. An EGC method was used to combine the output of $N$ branches to provide the receiving diversity. Besides, we studied the SKR performance of a post-selection based CV-QKD protocol using CD-Kennedy receiver with EGC in turbulent channels and compare the SKR performance with the protocol using homodyne receiver. Moreover, we proposed an equivalent transmittance method to facilitate the calculation of both the BER and SKR and used a Fenton-Wilkinson method to approximate the PDF of the equivalent transmittance. Numerical results demonstrated that the CD-Kennedy receiver can outperform the homodyne receiver in turbulent channels in terms of both BER and SKR performance. However, we found that BER and SKR performance advantage of CD-Kennedy receiver over homodyne receiver demonstrate opposite trends as the average transmittance increases, which indicates that two separate system settings should be employed for communication and key distribution purposes. Our work sheds a light on the development of practical CV-QKD system over satellite-ground links with atmospheric turbulence.

\bibliographystyle{IEEEtran}
\bibliography{egbib}

\appendices

\section{Derivation of $\mu_{eq}$ and $\sigma_{eq}^2$}\label{Append_A}

The $n$th order moment of a log-normal distributed variable $x \sim LN(\mu,\sigma^2)$ can be obtained as
\begin{equation}\label{n_order_moment}
\text{E}\left[x^n\right]=\exp\left(n\mu+n^2\sigma^2/2\right).
\end{equation}

Then we can rewrite the two constrains of Fenton-Wilkinson
method in \eqref{two_constrains} as
\begin{equation}\label{two_constrains_eq}
\begin{cases}
\text{E}[\eta_{eq}]=\sum_{i=1}^N \text{E}[\eta_{i}]\\
\text{E}[\eta_{eq}^2]=\text{E}\left[\left(\sum_{i=1}^N \eta_{i}\right)^2\right].
\end{cases}
\end{equation}

By using \eqref{n_order_moment}, we can rewrite the first equation in \eqref{two_constrains_eq} as
\begin{equation}\label{first_constrain}
\mu_{eq}+0.5\sigma_{eq}^2=\ln(N)+ \mu_0+0.5\sigma_0^2.
\end{equation}

Besides, the second equation in \eqref{two_constrains} can be further expressed as
\begin{equation}\label{second_constrain}
\begin{aligned}
\text{E}[\eta_{eq}^2]=\sum_{i=1}^N \text{E}\left[\eta_i^2\right]+2\sum_{i\neq j} \text{E}\left[\eta_i\eta_j\right].
\end{aligned}
\end{equation}

By using \eqref{n_order_moment}, we can obtain
\begin{equation}
\begin{cases}
\text{E}[\eta_{eq}^2]=\exp\left(2\mu_{eq}+2\sigma_{eq}^2\right)\\
\text{E}\left[\eta_i^2\right]=\exp\left(2\mu_{0}+2\sigma_{0}^2\right).
\end{cases}
\end{equation}

To calculate $\text{E}\left[\eta_i\eta_j\right]$, we define a new random variable
\begin{equation}
\eta_{ij}\triangleq \eta_i\eta_j
\end{equation}
\noindent for $i\neq j$. It is easy to verify that $\eta_{ij}$ is also a log-normal distributed variable with $\eta_{ij} \sim LN(2\mu_0,2(1+\rho)\sigma_0^2)$. Therefore, $\text{E}\left[\eta_i\eta_j\right]$ becomes the expectation of $\eta_{ij}$, i.e.,
\begin{equation}
\text{E}\left[\eta_i\eta_j\right]=\exp\left(2\mu_0+(1+\rho)\sigma_0^2\right).
\end{equation}

Then eq. \eqref{second_constrain} can be rewritten as
\begin{equation}\label{second_constrain_2}
\begin{aligned}
&\exp\left(2\mu_{eq}+2\sigma_{eq}^2\right)\\
&\quad =N\exp\left(2\mu_{0}+2\sigma_{0}^2\right)+N(N-1)\exp\left(2\mu_0+(1+\rho)\sigma_0^2\right).
\end{aligned}
\end{equation}

Then we can easily obtain $\mu_{eq}$ and $\sigma_{eq}^2$ as \eqref{mu_eq_and_sigma_eq_2} by combining \eqref{first_constrain} and \eqref{second_constrain_2}.

\section{Derivation of $\mu_{eq,hd}$ and $\sigma_{eq,hd}^2$}\label{Append_B}

Now we can define a new random variable
\begin{equation}
\eta_{sq,i}\triangleq \sqrt{\eta_i}
\end{equation}
\noindent for any $i=1,2,\cdots,N$.

It is easy to verify that  $\eta_{sq,i}$ is also a log-normal distributed variable with $\eta_{sq,i} \sim LN(0.5\mu_0,0.25\sigma_0^2)$. Besides, the correlation coefficient $\rho_{ij}$ between $\ln(\eta_{sq,i})=0.5 \ln(\eta_i)$ and $\ln(\eta_{sq,j})=0.5 \ln(\eta_j)$ when $i\neq j$ can be obtained as
\begin{equation}
\begin{aligned}
\rho_{ij}&\triangleq \frac{\text{E}[\ln(\eta_{sq,i})\ln(\eta_{sq,j})]-\text{E}[\ln(\eta_{sq,i})]\text{E}[\ln(\eta_{sq,j})]}{\sqrt{\text{Var}[\ln(\eta_{sq,i})]\text{Var}[\ln(\eta_{sq,j})]}}\\
&=\frac{0.25\text{E}[\ln(\eta_i)\ln(\eta_j)]-0.25\text{E}[\ln(\eta_i)]\text{E}[\ln(\eta_j)]}{0.25\sqrt{\text{Var}[\ln(\eta_{i})]\text{Var}[\ln(\eta_{j})]}}\\
&=\rho,
\end{aligned}
\end{equation}
\noindent where the last step is directly followed from \eqref{rho}.

Then for $\eta_{eq}^{hd}\triangleq \sum_{i=1}^N \sqrt{\eta_{i}}$, the two constrains of Fenton-Wilkinson method can be expressed as
\begin{equation}\label{two_constrains_hd}
\begin{cases}
\text{E}\left[\eta_{eq}^{hd}\right]=\sum_{i=1}^N \text{E}\left[\eta_{sq,i}\right]\\
\text{E}\left[\left(\eta_{eq}^{hd}\right)^2\right]=\text{E}\left[\left(\sum_{i=1}^N \eta_{sq,i}\right)^2\right].
\end{cases}
\end{equation}

By comparing \eqref{two_constrains_hd} and \eqref{two_constrains_eq}, we can find that $\mu_{eq,hd}$ and $\sigma_{eq,hd}^2$ can be obtained by simply replacing $\{\mu_0,\sigma_0^2\}$ with $\{0.5\mu_0,0.25\sigma_0^2\}$ in \eqref{mu_eq_and_sigma_eq_2}, which results in \eqref{mu_eq_and_sigma_eq_2_homodyne}.

\end{document}